\documentclass[prx,aps,twocolumn,superscriptaddress]{revtex4-1}

\usepackage{epsfig}
\usepackage{natbib}
\usepackage{mathrsfs}
\usepackage{amsmath}

\begin{document}

\title{Mermin-Wagner physics, $(H,T)$ phase diagram, and candidate quantum
spin-liquid phase in the spin-1/2 triangular-lattice antiferromagnet
Ba$_8$CoNb$_6$O$_{24}$}

\author{Y. Cui}
\affiliation{Department of Physics and Beijing Key Laboratory of
Opto-electronic Functional Materials $\&$ Micro-nano Devices, Renmin
University of China, Beijing, 100872, China}

\author{J. Dai}
\affiliation{Department of Physics and Beijing Key Laboratory of
Opto-electronic Functional Materials $\&$ Micro-nano Devices, Renmin
University of China, Beijing, 100872, China}

\author{P. Zhou}
\affiliation{Department of Physics and Beijing Key Laboratory of
Opto-electronic Functional Materials $\&$ Micro-nano Devices, Renmin
University of China, Beijing, 100872, China}

\author{P. S. Wang}
\affiliation{Department of Physics and Beijing Key Laboratory of
Opto-electronic Functional Materials $\&$ Micro-nano Devices, Renmin
University of China, Beijing, 100872, China}

\author{T. R. Li}
\affiliation{Department of Physics and Beijing Key Laboratory of
Opto-electronic Functional Materials $\&$ Micro-nano Devices, Renmin
University of China, Beijing, 100872, China}

\author{W. H. Song}
\affiliation{Department of Physics and Beijing Key Laboratory of
Opto-electronic Functional Materials $\&$ Micro-nano Devices, Renmin
University of China, Beijing, 100872, China}

\author{J. C. Wang}
\affiliation{Department of Physics and Beijing Key Laboratory of
Opto-electronic Functional Materials $\&$ Micro-nano Devices, Renmin
University of China, Beijing, 100872, China}

\author{L. Ma}
\affiliation{High Magnetic Field Laboratory, Chinese Academy of Sciences,
Hefei 230031, China}

\author{Z. Zhang}
\affiliation{State Key Laboratory of Surface Physics and Laboratory of
Advanced Materials, Department of Physics, Fudan University, Shanghai
200433, China}

\author{S. Y. Li}
\affiliation{State Key Laboratory of Surface Physics and Laboratory of
Advanced Materials, Department of Physics, Fudan University, Shanghai
200433, China}
\affiliation{Collaborative Innovation Center of Advanced Microstructures,
Nanjing 210093, China}

\author{G. M. Luke}
\affiliation{Department of Physics and Astronomy, McMaster University,
Hamilton L8S 4M1, Canada}
\affiliation{Canadian Institute for Advanced Research, Toronto M5G 1Z8, Canada}

\author{B. Normand}
\affiliation{Neutrons and Muons Research Division, Paul Scherrer Institute,
CH-5232 Villigen-PSI, Switzerland}

\author{T. Xiang}
\affiliation{Institute of Physics, Chinese Academy of Sciences, Beijing 100190,
China}
\affiliation{Collaborative Innovation Center of Quantum Matter, Beijing 100190,
China}

\author{W. Yu}
\email{wqyu\_phy@ruc.edu.cn}
\affiliation{Department of Physics and Beijing Key Laboratory of
Opto-electronic Functional Materials $\&$ Micro-nano Devices, Renmin
University of China, Beijing, 100872, China}



\begin{abstract}

Ba$_8$CoNb$_6$O$_{24}$ presents a system whose Co$^{2+}$ ions have an effective
spin 1/2 and construct a regular triangular-lattice antiferromagnet (TLAFM)
with a very large interlayer spacing, ensuring purely two-dimensional
character. We exploit this ideal realization to perform a detailed
experimental analysis of the $S = 1/2$ TLAFM, which is one of the keystone
models in frustrated quantum magnetism. We find strong low-energy spin
fluctuations and no magnetic ordering, but a diverging correlation length
down to 0.1 K, indicating a Mermin-Wagner trend towards zero-temperature
order. Below 0.1 K, however, our low-field measurements show an unexpected
magnetically disordered state, which is a candidate quantum spin liquid.
We establish the $(H,T)$ phase diagram, mapping in detail the quantum
fluctuation corrections to the available theoretical analysis. These include
a strong upshift in field of the maximum ordering temperature, qualitative
changes to both low- and high-field phase boundaries, and an ordered regime
apparently dominated by the collinear ``up-up-down'' state.
Ba$_8$CoNb$_6$O$_{24}$ therefore offers fresh input for the development of
theoretical approaches to the field-induced quantum phase transitions of
the $S = 1/2$ Heisenberg TLAFM.

\end{abstract}

\maketitle

\section{Introduction}

The challenge of frustrated quantum magnetism now extends from theory and
numerics through experiment to materials synthesis. This challenge is to
characterize and to understand the effects of quantum spin fluctuations in
dimensions greater than 1. In two dimensions (2D), where the $S = 1/2$
square-lattice antiferromagnet (SLAFM) with Heisenberg interactions has clear
magnetic order with a suppressed moment ($m_s \simeq 0.61 m_0$, where $m_0$
is the full moment) and the kagome-lattice AFM has no order at all, the
triangular-lattice antiferromagnet (TLAFM) lies close to the boundary where
the frustration-driven quantum fluctuations are sufficient to destroy
$m_s$. While this situation has led to a range of exotic proposals for the
ground state of the Heisenberg TLAFM \cite{Anderson_1973,Momoi_JPSJ_1992,
Chernyshev_PRL_2006_97,Starykh_RPP_2015_78}, detailed studies have demonstrated
that the true ground state does in fact have a finite semiclassical magnetic
order, in a noncollinear $120^{\circ}$ structure \cite{Bernu_PRL_1992_69,
Capriotti_PRL_1999_82}, with a best estimate for $m_s$ of $0.41 m_0$
\cite{White_PRL_2007}.

Nevertheless, the strong frustration of the TLAFM leads to an extensive
renormalization of physical properties at all energy scales. As a consequence,
the dynamical and thermodynamic properties of the TLAFM have remained as a
long-standing conundrum due to the inadequacy of theoretical approximations,
the limitations of numerical approaches (including small system sizes in exact
diagonalization, the minus-sign problem in quantum Monte Carlo, and the 1D
restriction on density-matrix renormalization-group (DMRG) methods), and the
absence of pure 2D systems for experimental investigation. A full understanding
of the TLAFM would also aid the understanding of other exotic quantum states,
most notably spin liquids \cite{Balents_nature_2010}, unconventional
superconductors \cite{Norman_2013}, and systems with complex magnetic order
\cite{Yamamoto_PRL_2014_112}, in all of which geometric frustration has an
essential role.

Purely 2D models such as the TLAFM are difficult to realize in the 3D world,
and a further complication to experiment is that their physics is controlled
by the Mermin-Wagner theorem \cite{Mermin_PRL__1966_17}, which describes the
dominant effects of additional thermal fluctuations in the restricted phase
space of a low-dimensional system. Specifically, the theorem dictates that in
2D a continuous symmetry can be broken, allowing a finite order parameter,
only at exactly zero temperature. In practice, most experimental systems are
subject to a weak 3D coupling that stabilizes their semiclassical order, and
so examples of ``Mermin-Wagner order,'' meaning incipient order as $T
\rightarrow 0$, are rare.

Among the known compounds approximating the spin-1/2 TLAFM
\cite{Coldea_PRL__2001_86,Shimizu_PRL__2003_91}, the materials
Ba$_3$CoSb$_2$O$_9$ and Ba$_3$CoNb$_2$O$_9$ have attracted particular attention.
They have perfectly regular lattices of Co$^{2+}$ ions and preserve inversion
symmetry close to the plane, so that any Dzyaloshinskii-Moriya interactions
are too weak to break the continuous symmetry \cite{Shirata_PRL_2012,
Zhou_PRL_2012_109,Susuki_PRL__2013,Zhou_PRB_2014_89,Yokota_PRB__2014_90}.
The exchange interactions are found to be close to the pure Heisenberg form,
albeit with a weak anisotropy ($J_x = J_y \neq J_z$). Magnetic order occurs
at finite temperatures due to weak interplane coupling, which releases the
stricture of the 2D Mermin-Wagner theorem. The rich variety of competing
magnetically ordered phases at finite applied magnetic fields
\cite{Zhou_PRB_2014_89,Yamamoto_PRL_2014_112} provides evidence of
the expected frustration effects. Available theoretical approaches
\cite{Singh_PRL_2006_12, Chubukov_PRL__1994_72, Mourigal_PRB__2013_88,
Chubukov_PRB__2006_74, Singh_NJP_2012_14} suggest the presence of low-lying
and weakly dispersive excitations, which were observed only recently in
Ba$_3$CoSb$_2$O$_9$ \cite{Ma_PRL_116}, and whose effects on the thermodynamic
properties of the TLAFM lie beyond the semiclassical predictions of the
nonlinear $\sigma$ model \cite{Singh_PRL_1993_71, Singh_PRB__2006_74,
Singh_NJP_2012_14}.

Here we take an experimental approach to the physical properties of
the spin-1/2 TLAFM close to the Heisenberg point, which hinges on the
remarkable properties of the dielectric compound Ba$_8$CoNb$_6$O$_{24}$.
This system has regular triangular layers of effectively low-spin Co$^{2+}$
ions separated by a very large interlayer spacing, $c \simeq 18.9$ \r{A}
\cite{Mallinson_Angew_2005_117}, which ensures that, from a magnetic point
of view, it is ideally 2D. It has been argued very recently by analyzing
bulk thermodynamic measurements that the magnetic Hamiltonian is that of
an ideal TLAFM with only Heisenberg interactions \cite{Rawl_PRB_95}. Here
we combine magnetization, $g$-factor, susceptibility, specific-heat, and
$^{93}$Nb nuclear quadrupole and magnetic resonance (NQR and NMR) measurements
down to temperatures of 0.028 K, to show that above 0.1 K Ba$_8$CoNb$_6$O$_{24}$
provides a nearly ideal experimental illustration of textbook Mermin-Wagner
physics in a 2D magnetic system. However, below 0.1 K our NQR and NMR data
reveal an anomalous suppression of low-energy spin fluctuations, breaking
the trend towards zero-temperature magnetic order. Instead they indicate a
disordered phase, possibly a quantum spin liquid (QSL), at zero field, and
at finite fields and temperatures an anomalously broad regime of apparent
collinear ``up-up-down'' polarization.

The structure of this paper is as follows. In Sec.~II we present the material
and experimental methods, with which we first establish that the system has
an effective spin $S = 1/2$ and a nearest-neighbor AFM exchange coupling
$J \simeq 1.66$ K. In Sec.~III we show our results for the low-temperature
($T > 0.08$ K) susceptibility and specific heat, and illustrate briefly how
these thermodynamic quantities provide a semi-quantitative validation of
available theoretical (series-expansion and Schwinger-boson) approaches
to the TLAFM. Section IVA presents our NQR data, which confirm the lack of
spontaneous magnetization down to 0.028 K at zero field, but with a correlation
length that increases steeply upon cooling to 0.1 K, and then quantify the
departures from Mermin-Wagner behavior below this temperature. In Sec.~IVB we
show our NMR results for the spectrum and spin-lattice relaxation rate, which
we compile as a field-temperature phase diagram. In Sec.~V we discuss the
interpretation of these results in terms of quantum corrections to the
classical phase diagram, accompanied by the appearance of the unexpected
candidate QSL phase at low field and temperature. In Sec.~VI we provide a
short summary.

\begin{figure*}
\includegraphics[width=17cm,height=11cm]{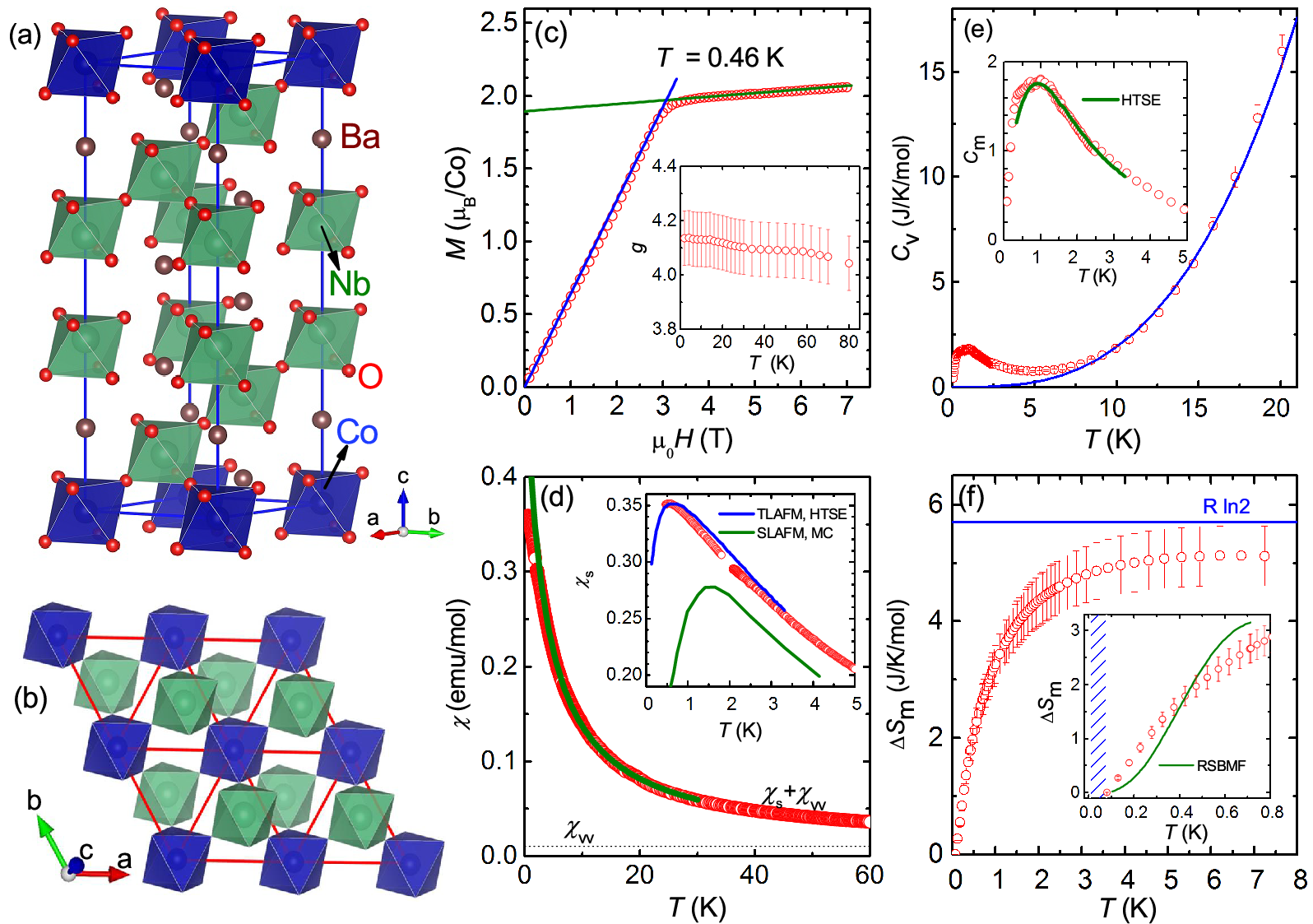}
\caption{\label{1} {\bf Lattice structure and magnetic properties of
Ba$_8$CoNb$_6$O$_{24}$.} (a) The triangular-lattice planes formed from the
CoO$_6$ units are separated by many nonmagnetic units, specifically 8 Ba$^{2+}$
and 6 NbO$_6$ layers. (b) A plane of CoO$_6$ octahedra, showing their
triangular-lattice configuration and the neighboring, corner-sharing NbO$_6$
octahedra that mediate the antiferromagnetic interactions. (c) Magnetization,
$M$; blue and green straight lines represent linear fits to the low- and
high-field data, whose origins lie respectively in the effective-spin and
van Vleck paramagnetic contributions. Inset: $g$-factor as a function of
temperature. (d) dc susceptibility measured under a field of 0.1 T. The solid
green line is a fit to a Curie-Weiss form, $\chi_{s}(T) = a/(T + \theta)$,
combined with an offset of $\chi_{vv}$, the constant van Vleck contribution
determined from $M$. The fitting parameters for the low-temperature regime
are $a = 1.7$ emu K/mol and $\theta = 3.5 \pm 0.5$ K. Inset: low-temperature
spin susceptibility; the blue line shows HTSE results for the Heisenberg
TLAFM, adapted from Ref.~\cite{Singh_PRL_1993_71}; the green line shows
Monte Carlo results for the Heisenberg SLAFM, adapted from
Ref.~\cite{Makivic_PRB__1991_43}. Both curves use the parameters $J = 1.66$ K
and $g = 4.13$. (e) Specific heat at zero field. The solid line is a fit to
$C_v = b T^3$. Inset: low-temperature magnetic specific heat, $C_m$, after
subtracting the phonon contribution; the green line is the HTSE result for
the Heisenberg TLAFM, adapted from Ref.~\cite{Singh_PRL_1993_71} with $J =
1.66$ K. (f) Magnetic entropy, $S_m(T)$, obtained by integrating the
specific-heat data above 0.08 K. The solid line shows the high-temperature
limit, $S_m(\infty) = {\rm R} \ln(2S + 1)$ in a spin-$S$ system, for $S =
1/2$. Inset: low-temperature magnetic entropy, showing for comparison
RSBMF results (see text) adapted from Ref.~\cite{Singh_NJP_2012_14} with
$J = 1.66$ K. Blue shading represents the temperature region excluded from
our analysis.}
\end{figure*}

\section{Material and Methods}

Polycrystalline Ba$_8$CoNb$_6$O$_{24}$ samples were synthesized by a
solid-state reaction method \cite{Mallinson_Angew_2005_117}. For structural
characterization we performed powder x-ray diffraction measurements and made
a complete Rietveld refinement using the Fullprof package. The material
crystallizes in the space group $P\bar3m1$, illustrated in Fig.~\ref{1}(a).
Co$^{2+}$ ions in CoO$_6$ octahedra form a corner-sharing geometry with
NbO$_6$ octahedra, constructing perfect layers of regular triangular lattices
[Fig.~\ref{1}(b)], with neighboring Co$^{2+}$ planes separated by eight Ba$^{2+}$
and six NbO$_6$ layers. The lattice parameters \cite{Mallinson_Angew_2005_117}
are $a = 5.789813$ \r{A}, which is almost identical to Ba$_3$CoNb$_2$O$_9$
($a = 5.7737$ \r{A}), but $c = 18.89355$ \r{A}, which is approximately three
times longer.

Our comprehensive experimental analysis of Ba$_8$CoNb$_6$O$_{24}$ combines
magnetization, $g$-factor, susceptibility, specific-heat, NQR, and NMR
measurements. Magnetization and susceptibility data were measured in a
PPMS-VSM for temperatures $T > 2$ K and in a $^3$He SQUID system for $0.6 \;
{\rm K} < T < 1.8$ K. The temperature-dependent $g$-factors were obtained
by field-sweep electron spin resonance (ESR) at a fixed frequency 9.397 GHz.
The specific heat was measured in a PPMS dilution refrigerator (DR), which
reached temperatures down to 0.08 K. The $^{93}$Nb ($I = 9/2$) NQR signal was
detected by the spin-echo technique in a DR system reaching temperatures
down to 0.028 K. Temperatures in both NQR and NMR were read from a RuO$_2$
thermometer and verified using the $^{63}$Cu NMR echo intensity of the coil,
which is inversely proportional to the temperature. The NMR spin-lattice
relaxation rate, $1/^{93}T_1$, was determined by the magnetization
inversion-recovery method, with the spin recovery, which showed no stretching
behavior, fitted by the standard $I = 9/2$ functions $m(t) = m(\infty) - a
[0.121 e^{-3t/T_1} + 0.56 e^{-10t/T_1} + 0.297 e^{-21t/T_1} + 0.022 e^{-36t/T_1}]$ for
NQR and $m(t) = m(\infty) - a [0.152 e^{-t/T_1} + 0.14 e^{-6t/T_1} + 0.153
e^{-15t/T_1} + 0.192 e^{-28t/T_1} + 0.363 e^{-45t/T_1}]$ for the NMR lines
\cite{MacLaughlin_PRB__1971_4,Fujita1984}. Theoretical values for the
thermodynamic quantities $\chi(T)$, $C_m(T)$, and $S_m(T)$ were digitized
from the cited literature and scaled appropriately.

It is necessary first to establish the effective spin of the Co$^{2+}$ ions in
Ba$_8$CoNb$_6$O$_{24}$. Figure \ref{1}(c) shows the magnetization, $M$, as a
function of field at a fixed temperature $T = 0.46$ K. $M$ increases rapidly,
from exactly zero field and completely linearly, up to a field $H_s = 3.00 \pm
0.04$ T; at this temperature there is no evidence for a feature at 1/3
magnetization, to which we return in Sec.~IV. Beyond $H_s$, a weak linear
increase is also observable, which can be ascribed to van Vleck paramagnetism
\cite{Zhou_PRB_2014_89}. $H_s$ is the saturation field required to polarize
fully the magnetic moment of Co$^{2+}$, as also measured for Ba$_3$CoSb$_2$O$_9$
\cite{Shirata_PRL_2012,Zhou_PRL_2012_109,Susuki_PRL__2013} and
Ba$_3$CoNb$_2$O$_9$ \cite{Zhou_PRB_2014_89,Yokota_PRB__2014_90}. The
saturation moment of the Co$^{2+}$ ions deduced from Fig.~\ref{1}(c) is
$m_s = 1.92 \pm 0.1 \mu_B$. ESR measurements of the $g$-factor at different
temperatures, shown in the inset of Fig.~\ref{1}(c), approach a constant
value, $g = 4.13 \pm 0.1$, below 20 K. The effective spin of the Co$^{2+}$
ions is therefore $m_s/g\mu_B \simeq 0.464$, which is consistent with
spin-$1/2$. This is the result expected from the crystal-field analysis
for Co$^{2+}$ in an octahedral environment with a weak trigonal distortion
\cite{Shirata_PRL_2012,Shiba_JPSJ_2003_72}, where the relatively strong
spin-orbit coupling leads to the formation of six nondegenerate Kramers
doublets and hence to an effective spin $S = 1/2$ with a large $g$-factor
at low temperatures.

The magnetic exchange coupling can be estimated from the same data. From
the corner-sharing geometry of neighboring CoO$_6$ and NbO$_6$ octahedra,
the dominant magnetic interactions between in-plane Co$^{2+}$ spins occur by
Co-O-O-Co and Co-O-Nb-O-Co superexchange couplings \cite{Zhou_PRB_2014_89};
the very long paths make the interaction strength extremely sensitive to
geometrical details and should preclude all but nearest-neighbor couplings.
For the effective $S = 1/2$ Co$^{2+}$ ions, one expects an XXZ spin model of
the form $H = \sum_{\langle ij \rangle} J_x (S_i^x S_j^x + S_i^y S_j^y) + J_zS_i^z
S_j^z$, where $\langle ij \rangle$ denotes only nearest-neighbor spins
\cite{Shirata_PRL_2012,Zhou_PRL_2012_109,Susuki_PRL__2013,Zhou_PRB_2014_89,
Yokota_PRB__2014_90}. In Ba$_3$CoSb$_2$O$_9$, the exchange coupling was stated
initially to be nearly isotropic (Heisenberg), with $J_x = J_y \approx J_z
\approx$ 18.2 K \cite{Shirata_PRL_2012,Zhou_PRL_2012_109,Susuki_PRL__2013},
but has since been found to have easy-plane character ($J_z/J_x \simeq 0.89$)
\cite{Ma_PRL_116}. In Ba$_3$CoNb$_2$O$_9$, double magnetic transitions at
1.36 K and 1.10 K in zero field indicate a weak easy-axis anisotropy, $J_z
 > J_x$ \cite{Zhou_PRB_2014_89,Yokota_PRB__2014_90,Matsubara_JPS__1982_51,
Miyashita_JPSJ_1985_54}, with $J_x$ of order 2 K.

Because Ba$_8$CoNb$_6$O$_{24}$ and Ba$_3$CoNb$_2$O$_{9}$ have an almost
identical planar structure, very similar exchange couplings are expected.
It is tempting to estimate the exchange directly from the magnetization
[Fig.~\ref{1}(c)], which by using the result $g \mu_B H_s = 4.5 J$ for the
Heisenberg TLAFM at $T = 0$ \cite{Yamamoto_PRL_2014_112} yields $J = 1.84
\pm 0.10$ K. However, this procedure has the weakness that our measurement
of $M$ is not made at zero temperature, which turns out to be a source of
significant inaccuracy due to the complex low-temperature physics of
Ba$_8$CoNb$_6$O$_{24}$. We use instead the result from our NMR measurements at
0.028 K, $H_s = 2.7 \pm 0.1$ T (Sec.~IV), which yields $J = 1.66 \pm 0.06$ K.
This result is fully consistent with the very recent estimate of
Ref.~\cite{Rawl_PRB_95}, which was made on the basis of comparison
with thermodynamic properties (Sec.~III).

\section{Thermodynamic Properties}

The dc susceptibility, $\chi(T)$, of the sample, measured under a field
of 0.1 T, is shown in Fig.~\ref{1}(d). $\chi(T)$ increases monotonically on
cooling down to 1 K and can be separated into two contributions, $\chi(T) =
\chi_{vv} + \chi_{s}(T)$, where $\chi_{vv} \simeq 0.019\mu_B$/Co/T, or 0.0106
emu/mol, is the van Vleck paramagnetic contribution determined from the
magnetization [Fig.~\ref{1}(c)]. $\chi_{s}(T)$ follows an approximate
Curie-Weiss (CW) form, $\chi_{s}(T) = a/(T + \theta)$, which represents
the average contribution of the coupled Co$^{2+}$ spins. In the low-temperature
regime ($T < 30$ K), we obtain a Weiss constant $\theta = 3.5 \pm 0.5$ K. From
the constant of proportionality, $a = 1.7$ emu K/mol, we deduce an effective
moment $\mu_{\rm eff} = 3.69 \mu_B$, which is fully consistent with our
determination of the saturation moment, $m_s$ (Sec.~II), because of the
value of the $g$-factor. These results are quantitatively consistent
with those of Ref.~\cite{Rawl_PRB_95}, and also with the results of
Ref.~\cite{Zhou_PRB_2014_89} for Ba$_3$CoNb$_2$O$_9$. At low temperatures,
$T \lesssim J$, $\chi_s(T)$ is expected to fall below the CW form as the
spins become correlated, giving the characteristic broad maximum revealed by
numerical approaches to the $S = 1/2$ TLAFM. In the inset of Fig.~\ref{1}(d),
we compare our data with the results for $\chi_s(T)$ obtained by the
high-temperature series-expansion (HTSE) method applied to the Heisenberg
TLAFM, adapted from Ref.~\cite{Singh_PRL_1993_71} using the parameters
$J = 1.66$ K and $g = 4.13$. The HTSE result and experimental measurements
are quantitatively consistent.

We stress that this comparison is not a fit, because with $J$ and $g$ fixed
there are no free parameters. While a best fit to the peak position would
return a smaller $J$ value, as noted above we rely on our measurement
of the low-$T$ saturation field to determine $J$. The errors in the
measured susceptibility arise from the sample mass (3\%), the determination
of the $g$-factor (3\%), and the accuracy of the SQUID data from the VSM
(5\%). In this context, the fact that our data and HTSE agree within 2.5\%
at all temperatures constitutes full agreement.

We draw attention to the generic property of the TLAFM that the peak in
$\chi_s(T)$ occurs at the anomalously low temperature $T \approx 0.4 J$;
this direct consequence of frustration can be contrasted with the behavior
of the unfrustrated SLAFM, adapted from Ref.~\cite{Makivic_PRB__1991_43} and
also shown in the inset of Fig.~\ref{1}(d), where the peak appears at $T
\approx J$. HTSE is by nature an approach from high temperatures, which
reaches its limits at the unusually low temperatures of the $\chi_s$ peak
in the TLAFM, and its use requires careful choice of representative Pad\'e
approximants. Thus this degree of consistency offers a benchmark both for
the capabilities of HTSE and for the degree to which Ba$_8$CoNb$_6$O$_{24}$
offers an ideal 2D $S = 1/2$ Heisenberg TFAFM.

Further valuable thermodynamic information is provided by the specific heat,
$C_v(T)$, shown in Fig.~\ref{1}(e). The absence of any sharp peak or cusp
in $C_v$ suggests that magnetic ordering is absent to the lowest temperature
(80 mK) we can access in this measurement. Because $C_v$ falls rapidly from
20 K to 7.5 K, following an exact $T^3$ behavior, we use this to subtract
the presumed phonon contribution and isolate the magnetic specific heat,
$C_m(T)$. This procedure can be followed with a high degree of confidence
because the characteristic energy scales of the phonons are manifestly very
high compared to the magnon contributions in this system, which peak at 1 K.
However, we caution that even very small residual uncertainties may be
important in the entropy analysis below, and contribute to the error bars
we display. $C_m(T)$, shown in the inset of Fig.~\ref{1}(e), confirms the
absence of ordering features, is dominated by a broad peak at $T \approx
1.0$ K, and below 0.3 K falls rapidly towards $C_m = 0$. While its shape over
the available data range shows no evidence for an activated form, a detailed
inspection at the lowest temperatures sets an upper limit of approximately
0.05 K on any possible spin gap. Once again we compare our data with
the HTSE result \cite{Singh_PRL_1993_71} for the Heisenberg TLAFM with $J =
1.66$ K, finding near-perfect quantitative agreement over the available range
of the HTSE data (0.3 K $\le T \le$ 2.5 K). As for the susceptibility, the
peak in $C_m(T)$ lies at a value anomalously low compared with the energy
scale of the SLAFM, as deduced from QMC \cite{Makivic_PRB__1991_43} and HTSE
methods \cite{Wang_PRB_1991_44,Bernu_PRB__2001_63}; again this result
reflects directly the effects of frustration in suppressing the overall
energy scale of the magnetic excitations \cite{Singh_PRB__2006_74}.

The magnetic entropy, $S_m(T) = \int C_m/T \, dT$, which we calculate from
our $C_m(T)$ data by integrating above 0.08 K, is shown in Fig.~\ref{1}(f).
At 7.5 K, we estimate that $S_m(T)$ saturates $90 \pm 10$\% of its total
value, $S_m(\infty) = R \ln 2$ for a spin-1/2 system. If the system were to
order at the lowest temperatures ($T < 0.08$ K), one might expect some of
the entropy to remain unaccounted for. However, because of the errors
accumulated in the integration, including those from the phonon subtraction,
we cannot draw any meaningful conclusions about possible missing entropy from
this result, other than that it is small. If, on the other hand, the ground
state were to have a spin gap, again our entropy results are consistent with
a value below 0.05 K [Fig.~\ref{1}(f)].

What we find at our lowest temperatures is that $S_m(T)$ has a very
rapid initial increase from $T = 0.08$ K, with 33$\%$ of $S_m(\infty)$
recovered by $T = 0.3J$ [inset, Fig.~\ref{1}(f)]. Here we compare our
data to the reconstructed Schwinger-boson mean-field (RSBMF) approach
\cite{Singh_NJP_2012_14}, a modified Schwinger-boson technique that
accounts correctly for the number of physical spin states and thus is
designed to capture the key properties of the TLAFM in the low-temperature
regime. By comparison with our data, the RSBMF formalism provides
semi-quantitative accuracy over the temperature range illustrated
(which matches the authors' claim for the validity of the method). However,
the form of $S_m(T)$ at lower temperatures is not well described, a result
on which we comment briefly below. Nevertheless, we note for perspective
that the type of nonlinear-$\sigma$-model approach so effective for the
SLAFM recovers only 5$\%$ of the total entropy at $T = 0.3 J$ in the TLAFM
\cite{Singh_PRL_1993_71,Chubukov_JPCM_1994_6} and therefore appears incapable
of providing a suitable account of frustrated systems.

We conclude that, over the full range of temperatures covered by our
thermodynamic measurements, Ba$_8$CoNb$_6$O$_{24}$ behaves as a model 2D
$S = 1/2$ TLAFM. Concerning the Hamiltonian governing the behavior of
the effective $S = 1/2$ spins, by comparison with the analysis of
Ref.~\cite{Rawl_PRB_95}, our $\chi_s(T)$ and $C_m(T)$ data contain no
evidence for an XXZ anisotropy, of Ising or XY type. Thus within the error
bars of our thermodynamic measurements, the effective model realized by the
material is a Heisenberg TLAFM. At a more microscopic level, an effective
Heisenberg interaction implies that the trigonal distortion of the CoO$_6$
octahedra, denoted by $\delta$ in Refs.~\cite{Shirata_PRL_2012,
Shiba_JPSJ_2003_72}, should vanish. In the structural refinement performed
on the basis of our powder x-ray diffraction measurments, we indeed obtain
an excellent account of our data without any trigonal distortion, and thus
find no evidence for a finite $\delta$ parameter within the error bars of
the fitting process.

\begin{figure}
\includegraphics[width=8.5cm,height=10.8cm]{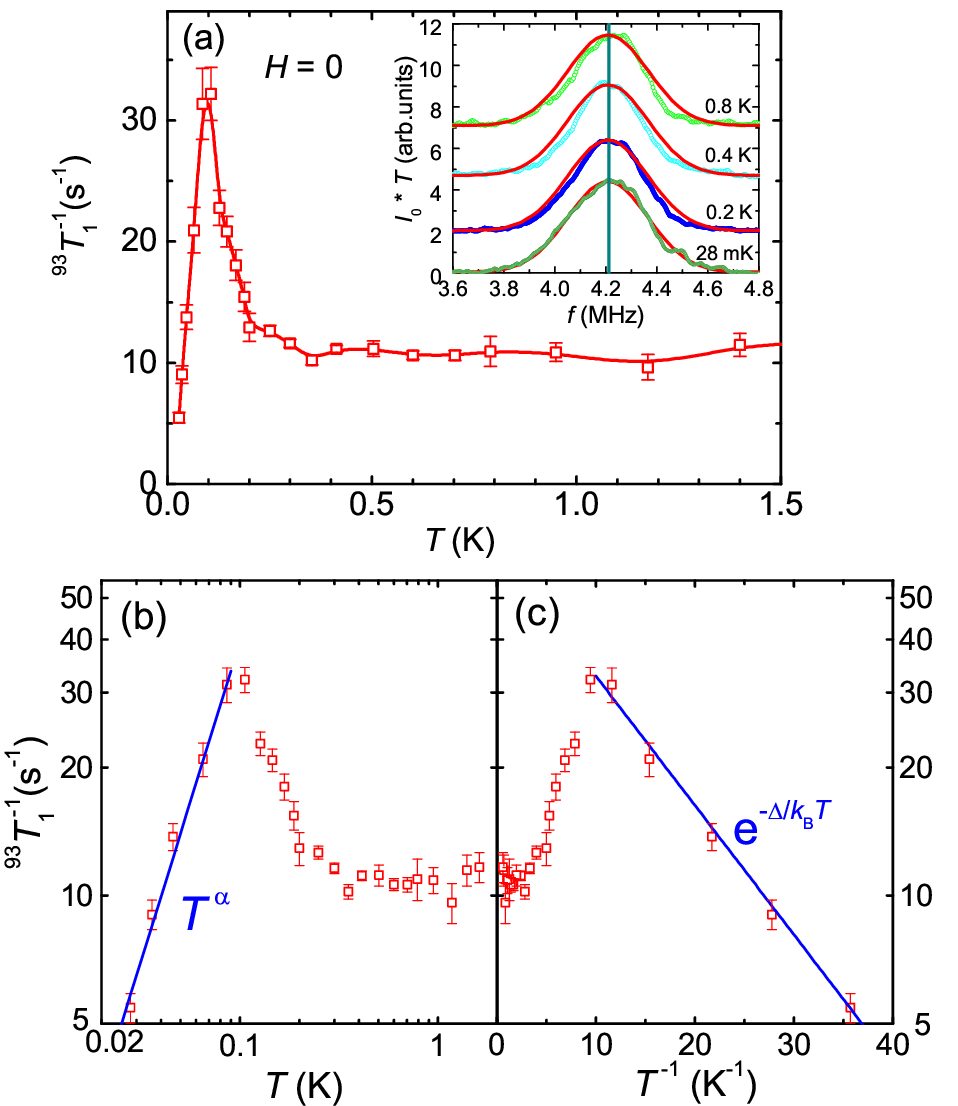}
\caption{\label{2} {\bf NQR measurements on Ba$_8$CoNb$_6$O$_{24}$.}
(a) Spin-lattice relaxation rate, $1/^{93}T_1$, measured at zero field as a
function of temperature. The upturn below 0.2 K indicates a progressive
increase of the correlation length upon cooling, which is cut off abruptly
below 0.1 K. Inset: zero-field $^{93}$Nb NQR spectra for temperatures from
0.8 K to 0.028 K, with the measured intensity multiplied by $T$. Red lines
represent the Gaussian fit to the spectrum at 0.028 K overlaid on all four
datasets. (b) $1/^{93}T_1$ data on logarithmic axes. The straight-line fit
(blue) indicates an approximate power-law form, $1/^{93}T_1 \propto T^\alpha$
with $\alpha = 1.51 \pm 0.06$. (c) $1/^{93}T_1$ data on semi-log axes as a
function of inverse temperature. The straight-line fit (blue) indicates an
approximate activated form with energy gap $\Delta = 0.070 \pm 0.003$ K.}
\end{figure}

\section{NQR and NMR Measurements}

\subsection{NQR and candidate QSL}

To probe both static and low-energy magnetic properties, we begin at zero
field by presenting low-temperature $^{93}$Nb NQR data. Here we report only
the signal with the shortest spin-lattice relaxation time, $^{93}T_1$, which
is over three orders of magnitude lower than the other times present and
arises in all probability from the Nb sites closest to the Co$^{2+}$ layers
(i.e.~with strong hyperfine coupling to the Co$^{2+}$ moments). The
$^{93}$Nb NQR spectra for excitations between $I_z = \pm 9/2$ and $I_z =
\pm 7/2$ are shown in the inset of Fig.~\ref{2}(a) for temperatures from 0.8
K down to 0.028 K and with their intensity, $I_0$, corrected for the Zeeman
factor, $1/T$. At $T = 0.028$ K, the spectrum is well fitted by a Gaussian
function (red line), and this fit can be applied also to the spectra at 0.2 K,
0.4 K, and 0.8 K. Clearly the spectra at all temperatures are centered on one
maximum, at the resonance frequency $f = 4.212$ MHz ($4 \nu_q$), and all have
the same echo intensity. The absence of any signal loss or NQR frequency
shift excludes completely the onset of any magnetic order down to 0.028 K.
Because the $^{93}$Nb nucleus is located directly above the center of one
triangle composed of three Co$^{2+}$ ions [Fig.~\ref{1}(b)], the static
hyperfine field at each $^{93}$Nb site should, if the Co$^{2+}$ moments order
with the $120^{\circ}$ coplanar pattern \cite{Yamamoto_PRL_2014_112}, have an
appreciable component normal to the TLAFM plane. Thus it is highly unlikely
that magnetic order could be missed in the NQR spectra.

The NQR spin-lattice relaxation rate shown in Fig.~\ref{2}(a) probes the
low-energy spin fluctuations. In general, $1/T_1 T = \sum_q A_{\rm hf}(q) \,
{\rm Im} \chi^{\pm} (q,\omega)/\omega|_{\omega\rightarrow 0}$, where $A_{\rm hf}$
is the hyperfine coupling constant and $\chi^{\pm}(q,\omega)$ the transverse
dynamic susceptibility. The fact that $1/^{93}T_1$ is of order 10 s$^{-1}$
indicates rather strong hyperfine coupling between the $^{93}$Nb nucleus and
the Co$^{2+}$ spins. $1/^{93}T_1$ also contains no evidence for long-range
order, but its upturn below 0.2 K indicates an increasing correlation
length, precisely as would be expected if the system approaches the
zero-temperature magnetic order anticipated by the Mermin-Wagner theorem
\cite{Mermin_PRL__1966_17}. To our knowledge, such a direct observation of
incipient ``Mermin-Wagner order'' in a purely 2D magnetic system has not
hitherto been obtained.

However, this textbook-quality Mermin-Wagner divergence is cut off quite
abruptly at 0.1 K, as Fig.~\ref{2}(a) shows clearly. Instead of continuing
to diverge as the temperature decreases, the $1/^{93}T_1$ signal shows a
sharp peak and a rapid drop to very low values. We stress again that the
temperature-normalized spectra are almost identical for all temperatures
[inset, Fig.~\ref{2}(a)], and that all of the magnetization recovery curves
we measure contain no sign of a stretched exponential form. As noted above,
these results provide no evidence for splitting or broadening of the peak,
as would occur in a finite-$T$ ordered phase. Similarly, they contain no
evidence for any spin-freezing, or other types of spin-glass behavior, at
such a low temperature, ruling out an inhomogeneous origin for the drop in
$1/^{93}T_1$. Despite the normalized echo intensities remaining constant for
all temperatures, a weak signal loss is observed close to 100~mK, which
reinforces the evidence for very strong low-energy spin fluctuations
coinciding with the peak in $1/^{93}T_1$. Thus these features appear to be
clear evidence for a transition or abrupt crossover to a zero-temperature
magnetically disordered phase, which is a candidate for an intrinsic QSL.

At the experimental level,
a QSL is typically analyzed by considering the temperature dependence of the
spin-lattice relaxation rate. An algebraic form indicates a gapless QSL while
an exponential (activated) form indicates a gapped QSL. Figure \ref{2}(b)
shows the data of Fig.~\ref{2}(a) on logarithmic axes, which suggest a
functional form $1/^{93}T_1 \propto T^\alpha$, i.e.~a gapless QSL with $\alpha
 = 1.51 \pm 0.06$. By contrast, Fig.~\ref{2}(c) shows the same data in the
activated form, which returns at least as good a fit to a gapped QSL with
$\Delta = 0.070 \pm 0.003$ K. We comment that the quoted errors are only
statistical, and do not include systematic errors. The only conclusion one
may draw is that it is impossible to make any meaningful deductions from a
fit to five data points covering half a decade in temperature. However, the
very low temperature scales involved make it impossible to broaden the fitting
window. Under the gapless scenario, such a small and half-integer exponent
would have to be the consequence of spin fractionalization. Under the gapped
scenario, such a small gap (of order $J/25$) would have few consequences at
all conventional experimental temperatures (and is very difficult to probe).
The only unambiguous conclusion from our NQR results is that new physics
sets in at an interaction energy scale of order 0.1 K. We defer a discussion
of possible theoretical scenarios to Sec.~V.

\begin{figure}
\includegraphics[width=8.5cm,height=11.2cm]{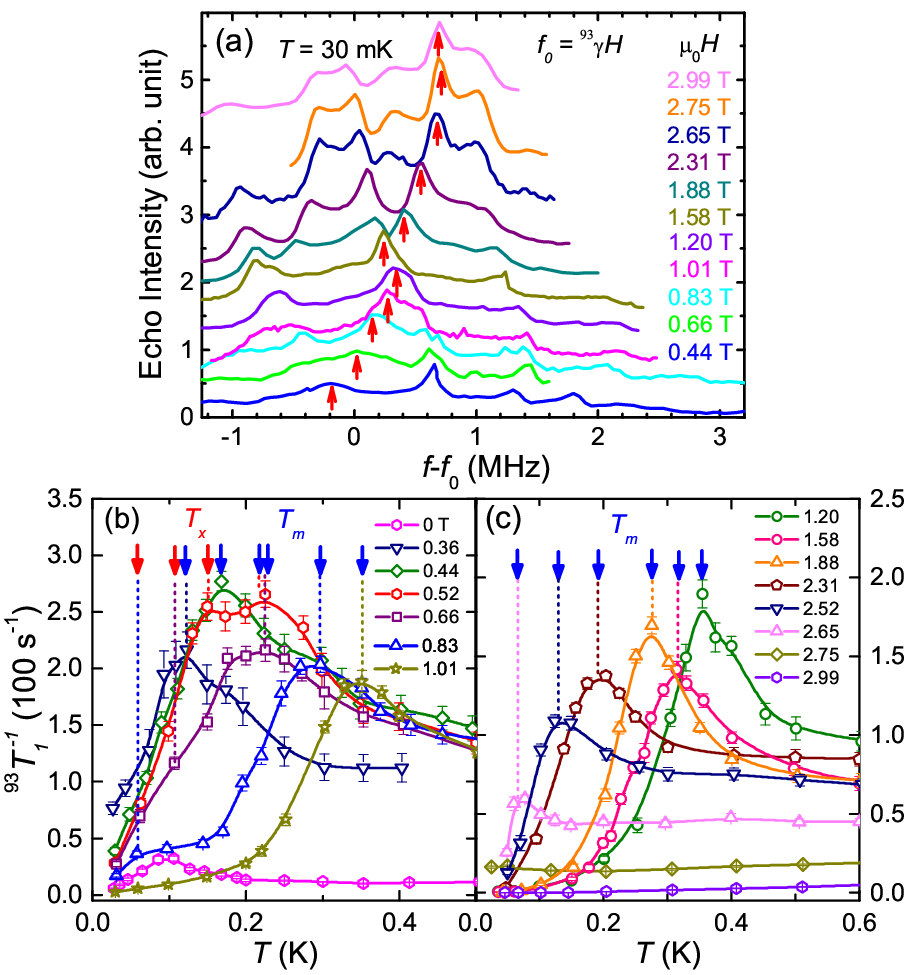}
\caption{\label{3} {\bf NMR response of Ba$_8$CoNb$_6$O$_{24}$ over a range
of applied fields.} (a) $^{93}$Nb field-sweep spectra displayed as a function
of frequency, $f - f_0$, at a fixed temperature $T = 30$ mK. The reference
frequency $f_0 = {^{93}\gamma}H$ is defined for each spectrum from the
measurement field, $H$, given in the legend. The spectra are offset vertically
for clarity. Arrows mark the spectral peak where the spin-lattice relaxation
rate, $1/^{93}T_1$, is measured. (b) $1/^{93}T_1$ shown as a function of
temperature for selected low fields. (c) $1/^{93}T_1 (T)$ for selected high
fields. Blue arrows mark the positions of the peaks we take to indicate the
magnetic transition. Red arrows for certain field values in panel (b) mark a
lower characteristic feature (see text).}
\end{figure}

\subsection{NMR and $(H,T)$ Phase Diagram}

To investigate the situation in more detail, we expand our studies of the
TLAFM to finite magnetic fields by performing NMR measurements at applied
fields up to and beyond saturation. Figure \ref{3}(a) shows $^{93}$Nb NMR
spectra taken over the full range of field values, all of which have a
dominant feature centered close to the frequency $f_0 = ^{93}\gamma H$,
where $^{93}{\gamma} = 10.421$ MHz/T is the gyromagnetic ratio of $^{93}$Nb,
accompanied by a number of other features spanning a broad frequency range.
We stress that the entire spectrum is the contribution of one single type
of $^{93}$Nb site, due to the $I = 9/2$ nuclear spin and the powder sample,
as we discuss next.

For the powder sample we have available, the NMR spectra show very strong
broadening due to the combination of quadrupolar corrections~\cite{Abragam},
anisotropic field effects, and the anisotropy of the hyperfine coupling in
the TLAFM~\cite{Brown2015}. While this makes a full assignment of the
different resonance frequencies to different field-orientation distributions
an impossible task, we begin our interpretation of the NMR data at high
fields, $H > H_s$. The spectrum at 2.99 T has a peak around $f - f_0
 = 0.7$ MHz, whose position [marked by the arrows in Fig.~\ref{3}(a)]
changes more rapidly with decreasing field than any other feature.
This indicates a strong hyperfine field with one specific orientation.
Although it is not possible to verify, this peak is consistent with an NMR
center line for field orientation $H {\parallel} ab$ for three reasons:
i) it is close to $f - f_0 = 0$; ii) for layered compounds, $H {\parallel} ab$
is the most probable orientation in a powder sample; iii) the hyperfine
coupling is highly anisotropic from $H {\parallel} ab$ to $H {\parallel} c$,
as observed for the Ba(2) sites in Ba$_3$CoSb$_2$O$_9$~\cite{Brown2015}, which
ensures a significant separation of contributions due to different crystallite
orientations. On this basis, the broad feature with frequencies from 0.2 to
1.2~MHz at 2.99~T can be ascribed to the combination of a distribution of
hyperfine fields (which vary linearly with the Larmor frequency, $\nu_L$)
and a distribution of second-order quadrupolar corrections (which vary as
$\nu^2_q/\nu_L$~\cite{Abragam}), both due to crystallites with orientations
close to $H {\parallel} ab$. By contrast, the broad peak from $-0.4$ to
0~MHz may be associated with field orientations close to $H \parallel c$,
where the hyperfine fields are weaker \cite{Brown2015}; however, these
assumptions should be verified by single-crystal NMR studies when possible.
Finally, the resonances visible in the spectra at $|f - f_0| \ge 1$~MHz
are due to satellite transitions, which in an $I = 9/2$ system are shifted
from the center line by $\pm {\nu}_q$, $\pm 2 {\nu}_q$, $\pm 3 {\nu}_q$, and
$\pm 4 {\nu}_q$~\cite{Abragam}.

To compare the spectra in a consistent manner, we focus only on the prominent
center peak of each one, which is marked by the arrows in Fig.~\ref{3}(a). As
the applied field is increased, the peak position shifts to a higher frequency,
which indicates an increase of the local hyperfine field. However, although the
shift $f - f_0$ is caused primarily by the hyperfine-field contribution at
applied fields above 1~T, it is contaminated strongly by the quadrupolar
corrections at low fields \cite{Abragam}.

Here we make two comments of relevance to the discussion below. First,
an integral featue of the finite-field phase diagram of the TLAFM is the
``up-up-down'' state, of two field-aligned and one anti-aligned spin(s) per
triangle. Our estimate of the line shift resulting from an up-up-down phase of
field-induced order in Ba$_8$CoNb$_6$O$_{24}$ is approximately 230~kHz. This
value is 1/3 of the shift observed in the fully polarized state ($H > H_s$)
and is the same at all Nb sites, with the result that there is no splitting.
However, the presence of such a shift is not well established in our data,
due not least to the growing quadrupolar corrections at lower fields.
Second, our NMR spectra contain no evidence for the mixing of Co and Nb
sites, i.e.~for an extrinisic source of disorder. While one may postulate
that nonmagnetic Nb impurities in the Co TLAFM, and magnetic off-plane Co
impurities, may have a significant effect on the properties of
Ba$_8$CoNb$_6$O$_{24}$, we are unable to identify any inequivalent Nb sites.
In the closely related material Ba$_3$CoNb$_2$O$_9$, two very sharp magnetic
ordering transitions are observed \cite{Zhou_PRB_2014_89}, which serves again
to underline how weak such disorder effects appear to be in the Co-Nb system.
Further, the magnetization recovery [$I(t)$] we measure at all fields and
temperatures fits very well to the standard function (Sec.~II), which again
reinforces the message of excellent sample quality.

In Figs.~\ref{3}(b) and \ref{3}(c) we show as functions of temperature the
NMR spin-lattice relaxation rates, also denoted $1/^{93}T_1$, measured at the
frequency of the spectral peak. A sharp maximum is clearly visible at every
applied field. The general trend is that the peak value of $1/^{93}T_1$
increases strongly with field up to $\mu_0 H \approx 0.5$ T, before decreasing
more slowly up to fields beyond $H_s$. We have verified (data not shown) that
such a relaxation peak is also present when measured at different frequencies
in all spectra, which indicates very strongly that this form of behavior is
intrinsic. The temperature at which the peak appears, marked by the arrows
in Figs.~\ref{3}(b) and \ref{3}(c), varies systematically with the field,
increasing to $T_m \approx 0.35$ K at $\mu_0 H = 1.20$ T before falling
monotonically to zero at $H = H_s$.

A pure Heisenberg model in an applied magnetic field retains a rotational
symmetry about the field axis. However, the Ba$_8$CoNb$_6$O$_{24}$ system
has crystalline anisotropies and spin-orbit coupling in addition to possible
experimental misalignments, as a result of which it is realistic to take the
field as an explicit breaking of the continuous spin symmetry. In this event,
the Mermin-Wagner theorem is no longer applicable and a real magnetic order
is induced, even in 2D, until it is suppressed by thermal fluctuations. Thus
we ascribe the peak in the NMR $1/^{93}T_1$ to a real magnetic ordering
transition occurring at temperature $T_m$, below which the ordered phases
have an anisotropy gap. The magnetic transition in Ba$_3$CoSb$_2$O$_9$ was
also reported on the basis of the peak in $1/T_1$~\cite{Brown2015}, although
in this material the order sets in at 3.8~K at zero field due to interlayer
(3D) coupling. At $H = 0.52$ T, we find a curious double-peak structure,
which by the same reasoning should be associated with a second characteristic
temperature scale. At $H = 0.88$ T, the lower temperature appears not as a
peak but as a second sharp drop, occurring around $T = 0.06$ K at the lower
end of a plateau in the relaxation rate. As one of these forms evolves into
the other, our ascription of a second temperature scale, $T_x$, to a weak
feature in the $\mu_0 H = 0.66$ T data is very tentative. At 1.01 T, the
feature and hence $T_x$ has clearly vanished.

\begin{figure}
\includegraphics[width=8.5cm,height=8.5cm]{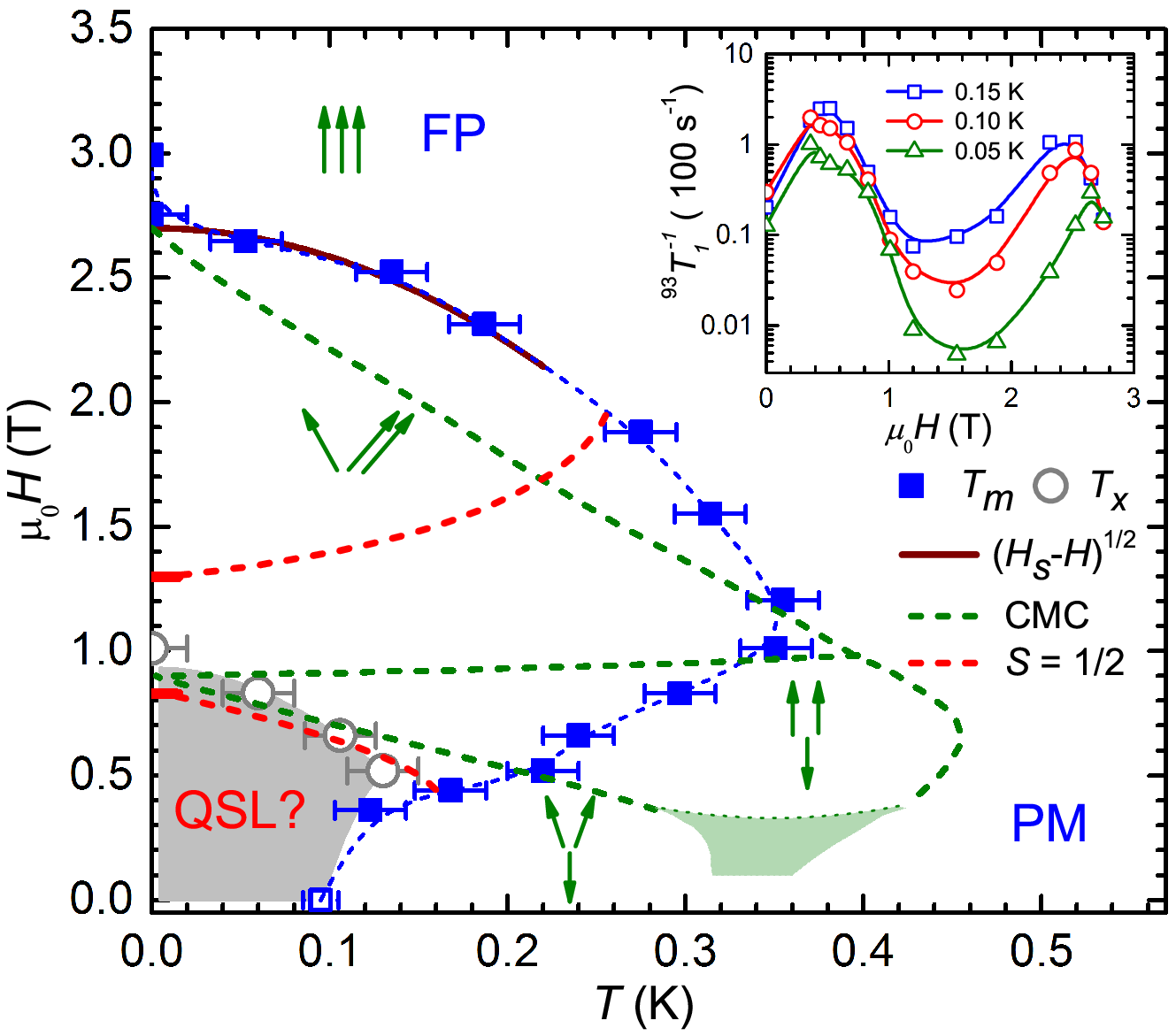}
\caption{\label{4} {\bf Field-temperature phase diagram of
Ba$_8$CoNb$_6$O$_{24}$.} $(H,T)$ phase diagram constructed from the spin-lattice
relaxation rate. Solid blue squares represent the temperature of the peak in
$1/^{93}T_1$, taken from Figs.~3(b) and 3(c), and open grey circles mark the
lower temperature scale found in $1/^{93}T_1$ at low fields [Fig.~3(b)] (see
text). The shaded region demarcates the candidate quantum spin liquid phase.
The solid line is a fit of the upper transition to the functional form $T_c(H)
 = c (H_s - H)^{0.5}$. Green arrow symbols and dashed lines represent the
phases and phase boundaries established by Monte Carlo simulations of the
classical TLAFM, adapted from Refs.~\cite{GMZ_JPCM_23} and
\cite{Seabra_PRB_84} and scaled to $J = 1.66$ K. Red ticks on the
field axis at $T = 0$ mark the boundaries of the up-up-down phase for
the $S = 1/2$ Heisenberg TLAFM, established by many authors and taken
in this case from Ref.~\cite{Farnell_JP_2009}. Red dashed lines mark
schematic finite-temperature phase boundaries anticipated for the
up-up-down phase in the $S = 1/2$ system, where it is favored by quantum
fluctuations. Inset: field dependence of $1/^{93}T_1$ shown at fixed
temperatures of 50, 100, and 150 mK. The presence of only two peaks
suggests that the entire intermediate regime accessible in experiment
may have only one type of field-induced order.}
\end{figure}

In Fig.~\ref{4}, we gather the $T_m$ and $T_x$ values determined from
$1/^{93}T_1$ in the form of an $(H,T)$ phase diagram. $T_m$ demarcates a
regime of field-induced magnetic order below $H_s \simeq 2.7$ T and a
maximum temperature of 0.35 K, while $T_x$ may be the upper limit in field
of the anomalous disordered region we find by NQR (Fig.~\ref{2}, marked
by the shaded region in Fig.~\ref{4}). As a guide to interpreting our
results, we have annotated Fig.~\ref{4} with a number of other lines.
Green dashed lines show the phase boundaries obtained by Monte Carlo
studies of classical (large-$S$) Heisenberg TLAFM \cite{GMZ_JPCM_23,
Seabra_PRB_84}, which obtain the four finite-$H$ phases represented by
the green arrows, namely distorted triangular, up-up-down, canted, and
fully polarized. The red bars on the field axis ($T = 0$) represent the
boundaries of the up-up-down plateau of the fully quantum $S = 1/2$
Heisenberg TLAFM, obtained by a number of methods and taken here from
Ref.~\cite{Farnell_JP_2009}. We stress two features of the $S = 1/2$
results, first that the same four phases obtained in the classical model
are present in the quantum case and second that, in contrast to the
classical case, the up-up-down plateau has a finite width at $T = 0$ as
a result of its stabilization by quantum fluctuations, which have a
generic preference for antiparallel spin configurations. Based on this
expectation, the red dashed lines sketch the possible phase boundaries
of the quantum up-up-down plateau at finite $T$. We stress that these lines
rank as no more than guides to the eye. There is as yet no theoretical or
numerical method that is capable of drawing the $(H,T)$ phase diagram of
the $S = 1/2$ TLAFM, which is a major reason why experimental results for
an almost-perfect TLAFM material such as Ba$_8$CoNb$_6$O$_{24}$ are so
important. The downward curvature of the lower boundary and upward curvature
of the upper one are based on the expectation that thermal fluctuations will
favor the collinear-spin up-up-down phase over the non-collinear distorted
triangular and canted phases (but the extent of this curvature is in essence
unknown). A very recent experiment has investigated the spin excitations of
the up-up-down state in Ba$_3$CoSb$_2$O$_9$ \cite{Kamiya_2017}, i.e.~in the
presence of finite 3D coupling and easy-plane anisotropy.

Because our measurements are performed on powder samples, it is very
difficult to determine the magnetic structure. In particular, the
difference between collinear and non-collinear ordered phases becomes
even more difficult to discern. In the inset of Fig.~\ref{4} is shown
$1/^{93}T_1$ as a function of field at the fixed temperatures $T = 50$,
100, and 150 mK. By the same criterion as in Figs.~\ref{3}(b) and \ref{3}(c),
these curves suggest that there are only two significant phase transitions
as the field is increased, and thus that the entire magnetically ordered
phase may be of only one type, rather than the anticipated three. We stress
for clarity that, despite the presence of the anomalous candidate QSL regime
at low $T$ and $H$ (shaded in Fig.~\ref{4}), one would still expect to
recover the behavior of the conventional $S = 1/2$ TLAFM for $T > 0.1$ K
(where thermal fluctuations overwhelm the additional low energy scale) and
for $H > 1$ T (where, empirically, the field suppresses quantum fluctuations
to a sufficient degree).

The possible origin of the candidate QSL is discussed in Sec.~VC. Outside
this regime, we can use our experimental measurements on Ba$_8$CoNb$_6$O$_{24}$
to benchmark the quantum phase diagram of the spin-1/2, 2D Heisenberg TLAFM
in a manner that has not been possible before. We identify four ways in
which quantum fluctuations act to revise the classical phase diagram.
(i) The field, $H_{\rm max}$, at which the maximum value of $T_m$ occurs
is 80\% higher in the $S = 1/2$ system than in the classical case.
(ii) In the classical Monte Carlo simulations, there are one or possibly
two finite-temperature phase transitions at very weak fields, but in the
$S = 1/2$ system this regime is entirely paramagnetic and the lower bound
on the ordered regime has a finite slope.
(iii) In the classical model, the phase boundary to the fully polarized state
approaches $H_s$ linearly, whereas in the $S = 1/2$ system the phase boundary
is fitted rather well, over approximately one decade of data in temperature,
by the form $T_m(H) \propto (H_s - H)^{1/2}$.
(iv) In the classical model one expects three different ordered phases under
the dome of $T_m(H)$, whereas in the experimental measurements there is no
evidence for a change of phase at finite temperatures.
We discuss these four points in detail in Sec.~VB.

\section{Discussion}

\subsection{Zero-field quantum physics: Mermin-Wagner}

The physics of a 2D quantum magnet with continuous symmetry is controlled
by the Mermin-Wagner theorem. While thermal fluctuations can act to reinforce
quantum fluctations in gapped systems, their effect on candidate ordered
phases is a systematic suppression. Thus in experiment one may observe only
an incipient order, which is best characterized by a correlation length,
$\xi$, or time, $\tau$, that increases with decreasing temperature towards a
divergence at $T = 0$. This is exactly the behavior we observe in the NQR
spin-lattice relaxation rate, shown in Fig.~\ref{2}(a), where the diverging
trend becomes clear below 0.2 K. However, the presence of the anomalous
disordered state at the lowest temperatures ($T < 0.1$ K) spoils a more
systematic characterization of this divergence.

The TLAFM is a keystone model in quantum magnetism because it offers one of
the basic realizations of a geometry that is frustrated for AFM interactions.
At zero field, where one may wish to investigate the consequences of
frustration for thermodynamic quantities, correlation functions, and the
dynamical response, its primary effect is the suppression of characteristic
energy scales. For a deeper understanding of the anomalous low energy scale
observed in the thermodynamic quantities of Fig.~\ref{1}, the most
sophisticated methods available for the Heisenberg TLAFM are HTSE
\cite{Singh_PRL_2006_12,Singh_PRB__2006_74}, self-consistent spin-wave
\cite{Chubukov_PRB__2006_74,Mourigal_PRB__2013_88}, and RSBMF treatments
\cite{Singh_NJP_2012_14}. These indicate that the one-triplet excitation
band is both extremely flat and unusually low-lying, with most of its weight
concentrated around $E \approx 0.6J$, similar to the recent observation in
Ba$_3$CoSb$_2$O$_9$ \cite{Ma_PRL_116}. This is a direct consequence of the
band being driven downward by the low-lying two-triplet excitation continuum,
which shows strong triplet interactions at all wave vectors due to the high
frustration. This behavior stands in sharp contrast to the
SLAFM, where the bands disperse uniformly up to energies $E \approx 2J$
\cite{Jarrell_PRL_1992_68}. The frustration-renormalized energy scale is
completely consistent with the temperature dependence we benchmark in the
peak features of $\chi(T)$ and $C_m(T)$. The unusual low-energy excitations
of the TLAFM have been variously described as ``roton-like'' or as evidence
of fermion deconfinement \cite{Singh_PRL_2006_12,Chubukov_PRB__2006_74}.

To our knowledge, Ba$_8$CoNb$_6$O$_{24}$ allows the first direct comparison
between theory and a real 2D material for the $S = 1/2$ Heisenberg TLAFM.
If one takes seriously the small but finite quantitative discrepancies
between the HTSE results \cite{Singh_PRL_1993_71} and our data for $C_m(T)$
[Fig.~\ref{1}(e)] and $\chi(T)$ [Fig.~\ref{1}(d)], one requires an explanation
for possible additional contributions on the low side of the peaks in both
quantities. The specific heat contains contributions from all types of
excitation, and appears to show unexpected weight around 0.5 K. The
susceptibility is sensitive only to finite-spin excitations, primarily
triplets ($S = 1$), and the peak in this quantity appears to lie at a
temperature 10-15\% lower than HTSE would forecast. Thus it is possible
that this method is not quite capturing the full extent of the suppression
of the one-triplet band due to the frustration of the TLAFM. Experimental
uncertainties notwithstanding, the RSBMF approach appears to underestimate
the low-temperature entropy [Fig.~\ref{1}(f)] and may not be capturing the
correct temperature dependence. One possible exotic explanation for this
result \cite{Trumper_PC} might be the incomplete binding of spinon degrees
of freedom contained within the slave bosons.

More generally, both results may be explained if the ``roton gap,'' the
effective bandwidth, the weight in the two-triplet sector, or any other
features affecting the density of states in the spin spectrum, all of
which require a correct accounting for quantum fluctuation effects, are not
reproduced perfectly by the theoretical or numerical approaches applied.
However, we caution that the small mismatch between our data and the
theoretical results shown in Fig.~\ref{1} cannot be interpreted unambiguously
as evidence for shortcomings in the theories, as it may be a consequence of
experimental uncertainties, particularly in the entropy, or of small
additional terms in the magnetic Hamiltonian. While it is clear from the
appearance of the candidate QSL phase (Sec.~VC) that such terms are indeed
present, we comment that energy scale for the QSL regime does not match
the temperatures at which the discrepancies in $\chi(T)$ and $C_m(T)$ appear,
although it could indeed be relevant for the interpretation of $S_m(T)$. A
notable candidate for such a term would be a weakly non-Heisenberg anisotropy
in the exchange couplings, which is already documented in Ba$_3$CoNb$_2$O$_{9}$,
although we are unable to obtain any evidence for this in our measurements. We
defer a further analysis of this point until single crystals become available.

\subsection{Quantum fluctuation corrections}

Here we discuss the differences between the classical and measured
(``quantum'') phase diagrams shown in Fig.~\ref{4}, retaining the
numbering scheme of Sec.~IVB.

\noindent
(i) The field at which the maximum of $T_m$ appears in the classical case,
$H_{\rm max} \approx 2J$ in dimensionless units, coincides with the center of
the most stable of the ordered phases, which is the up-up-down phase. The very
significant (approximately 80\%) rise of this optimum field in the quantum
system appears to reflect the much greater stability of the up-up-down phase
due to the strong quantum fluctuations of the $S = 1/2$ system. In fact our
choice of the position of the upper red dashed line in Fig.~\ref{4} was made
on the basis of an approximate symmetry criterion with respect to $H_{\rm max}$.

\noindent
(ii) In the classical Monte Carlo simulations, there is a problem at
zero field in that order is precluded by the Mermin-Wagner theorem but
is present at a finite temperature at any finite field. In the quantum case,
the lower bound on the ordered regime has a finite slope, which has a clear
interpretation in terms of the competition between the disordering effects
of thermal fluctuations and the ordering effects of a rising field. The slope
of this line, which is approximately linear if the QSL regime is ignored,
would be one topic for a more detailed analysis. We comment that the shaded
region in the classical Monte Carlo phase diagram denotes an uncertainty over
the nature of the thermal transition(s) that is reflected in a contrast
between the results of Refs.~\cite{GMZ_JPCM_23} and \cite{Seabra_PRB_84},
but that this appears to be a moot point in the quantum system.

\noindent
(iii) The contrasting functional forms of the magnetic phase boundary near
saturation, $T_m(H) \propto (H_s - H)^a$ with $a = 1$ in the classical system
and $a = 0.5$ in the quantum one, can be taken as a direct expression of the
qualitative differences resulting from quantum fluctuations. In this case
their effect is to confer extra stability on the field-induced ordered
phase, possibly changing its nature [point (iv) below], and to alter the
way in which this order is lost as thermal fluctuations increase.

\noindent
(iv) More mysterious is the nature of the field-induced ordered phase. As
noted in Sec.~IVB, three different ordered phases are expected between
$H = 0$ and $H = H_s$ in the classical model, and in the quantum model at
$T = 0$. Our experiments offer no evidence for a change of phase. Taking
this result at face value, it is possible that the thermal destabilization
of the non-collinear phases (distorted triangular and canted) in favor of
the collinear up-up-down phase is truly a strong effect, which is complete
below 50 mK in Ba$_8$CoNb$_6$O$_{24}$. However, as also discussed above, it
is difficult to exclude the possibility that the three ordered phases are
simply indistinguishable in the present powder experiments.

\subsection{Quantum disordered phase}

The most surprising feature of our results is the appearance of the
magnetically disordered phase at low temperature and low fields. Such a
phase is not expected in any studies of the $S = 1/2$ TLAFM at zero field
or zero temperature. Clearly the most basic question to address is whether
this phase is a consequence of intrinsic quantum physics, due to a weak
additional term in the magnetic Hamiltonian, or whether it could arise
from an extrinsic factor, such as sample disorder. While it is difficult
to make a definitive statement about this eventuality, the most obvious
type of randomness in Ba$_8$CoNb$_6$O$_{24}$ would be ``anti-site'' disorder,
namely non-magnetic Nb ions in the Co planes and magnetic Co ions on some
of the Nb sites. Nonmagnetic sites (vacancies) in an ordered 2D $S = 1/2$
magnet are found in theoretical studies \cite{Bulut_PRL_62} to cause a weak
reinforcement of order, at least if virtual electronic hopping to the site
is also prohibited. In the Ba$_8$CoNb$_6$O$_{24}$ geometry, a Co ion off the
plane might be expected to form a net singlet state with the three spins of
the neighboring in-plane triangle, creating an effective three-site vacancy,
and for the same reason this should not lead to a magnetically disordered
phase. In experiment, all types of disorder have rather clear fingerprints
in the NMR spectra, and as noted in Sec.~IVB we are not able to detect any
spectral features corresponding to Nb ions with a different type of
environment.

Another extrinsic factor would be the coupling of the electronic spins
to the nuclear-spin subsystem. In ordered magnetic materials, this coupling
can give rise to a shift, or ``pulling'' \cite{pulling63}, of the NMR
frequency and a significant literature has built up concerning the
measurement and quantitative description of this effect, particularly in
Mn-ion systems \cite{rdkpt99}. To estimate whether the interaction with
nuclear spins could be responsible for our observations, we first consider
the magnitude of the hyperfine coupling. The transferred hyperfine coupling
from the Co electronic spins to the Nb nuclear spins, which is the pathway
by which we probe the TLAFM in Ba$_8$CoNb$_6$O$_{24}$, has an interaction
constant $A_{\rm hf} = 0.04$ T per $\mu_B$. The associated energy scale is
calculated as $E_{\rm hf} = \gamma  A_{\rm hf} I S$, where $S = 1.9 \mu_B$ is
the magnetic moment per Co electronic spin and $I = 9/2$ the Nb nuclear
spin, whence $E_{\rm hf} = 0.14$ mK, which is much too small to be relevant
to our observations. However, there is also a direct hyperfine coupling of
the Co electronic to the Co nuclear spins ($I = 7/2$), which is approximately
$A_{\rm hf}' = 5$ T per $\mu_B$ in systems with local Co-ion coordination similar
to that in Ba$_8$CoNb$_6$O$_{24}$ \cite{Mukuda_JPSJ_06}, and the resulting
energy scale can be estimated as $E_{\rm hf}' = 0.016$ K. Once again, this
channel appears too small to be responsible for effects setting in at 0.1 K,
as we observe in Figs.~\ref{2}, \ref{3}, and \ref{4}.

Qualitatively, it is in any case not clear how the hyperfine interaction would
affect the electronic spin state in a thermally disordered magnetic phase, of
the type relevant here down to 0.1 K. Pulling requires a finite ordered moment
to produce the observed frequency shift \cite{Zaliznyak_JETP_96}. In the Ising
magnet LiHoF$_4$ \cite{RonnowSci05}, coupling to the nuclear spins acts to
extend the regime of magnetic order of the electronic spin system, which is
the opposite of the physics we observe. On general grounds, a coupling of the
electronic spin system to the very high density of low-energy excitations
introduced by the nuclear-spin system does not seem likely to establish a
situation where the density of low-energy spin excitations appears to vanish
towards zero energy. In one study of nuclear-spin coupling to a disordered
electronic spin system \cite{rtz16}, the magnetic spectrum of the composite
system remains gapless for systems with no nontrivial topological term, which
should be the relevant one in 2D, and does not appear to lose low-energy
weight. Although the coupled Hamiltonian belongs to the general class of
Kondo-type models, it is difficult to conceive of Kondo-type physics (local
singlet formation) for localized electrons interacting with $I = 7/2$ nuclear
spins. Thus, within the scope of the present study, we conclude that
nuclear-spin coupling cannot present a candidate origin for our observations.
We recall that, in an intrinsic state of the electronic spins alone, the
mutual interaction energy scale ($J = 1.66$ K) is two orders of magnitude
higher than the hyperfine coupling scale.

Turning to a possible intrinsic origin, we begin by reviewing other candidate
terms in the magnetic Hamiltonian. A 3D coupling is not expected on an energy
scale of 0.1 K in a system as 2D as Ba$_8$CoNb$_6$O$_{24}$, and in any case
this would promote finite-temperature order. Motivated by the material
Cs$_2$CuCl$_4$ \cite{Coldea_PRL__2001_86}, which shows QSL behavior in a
spatially anisotropic TLAFM, systems in which every triangle has one strong
and two weak bonds have been of enduring interest \cite{Chen_PRB_2013}.
However, there is no evidence in the NMR spectra for the onset of a
low-temperature lattice distortion, which would also be most unexpected at
such low energy scales.

Spin-orbit coupling has been studied intensively in insulating magnets
as the possible origin of chiral QSL states and Kitaev spin liquids. In
conventional local-moment systems, spin-orbit coupling is manifest in the
Dzyaloshinskii-Moriya interaction on chemical bonds lacking inversion
symmetry and in exchange anisotropies on inversion-symmetric bonds. For
the Co$^{2+}$ ions in Ba$_8$CoNb$_6$O$_{24}$, the spin-orbit energy scale,
$\lambda \approx 250$ K, is strong and the lowest Kramers doublet of the
crystal-field manifold defines the effective $S = 1/2$ degree of freedom,
along with a possible exchange anisotropy in the event of a finite trigonal
distortion \cite{Shirata_PRL_2012,Shiba_JPSJ_2003_72}. Within this effective
model, the possibility of effective Dzyaloshinskii-Moriya interactions is
excluded by the fact that the bonds of the Ba$_8$CoNb$_6$O$_{24}$ lattice
[Figs.~\ref{1}(a) and \ref{1}(b)] are inversion-symmetric (unless sample
disorder were to produce such terms around an impurity). Although an
effective exchange anisotropy is the strongest candidate for an additional
term in the spin Hamiltonian, we stress that XXZ physics has been found to
produce no abrupt or significant changes to the zero-temperature phase
diagram \cite{Yoshikawa_JPSJ_73}, and thus that such a term would not be
capable of creating a disordered magnetic state. Indeed, the fact that this
term breaks the continuous symmetry, in whole or in part, thereby lifting
the Mermin-Wagner constraint, would seem more likely to promote magnetic order
than to suppress it. While anisotropy of the single-ion type is not relevant
for $S = 1/2$ systems, the situation for the effective $S = 1/2$ entities
in Co$^{2+}$ is more complex. In this context it is worth noting that a
temperature-driven effective-low-spin to high-spin crossover appears to
occur in Ba$_3$CoNb$_2$O$_9$ between 100 and 200 K \cite{Zhou_PRB_2014_89},
i.e.~on the energy scale of $\lambda$, but it remains difficult to invoke
such behavior (which is governed by $d$-level crystal-field splittings on
100 meV energy scales) at 0.1 K.

Of the limited selection of remaining candidate mechanisms, a next-neighbor
coupling, $J_2$, has been shown \cite{Manuel_PRB_1999,Kaneko_JPSJ_2014,
Li_PRB_2015,Zhu_PRB_2015} to produce a QSL in the narrow range of interaction
strengths $0.06 \lesssim J_2/J \lesssim 0.17$. While all of the recent studies
agree rather closely on the boundaries of this regime, they are completely
divided on the nature of the QSL, with variational Monte Carlo (VMC) methods
suggesting a gapless state \cite{Kaneko_JPSJ_2014}, also supported by the use
of more sophisticated quantum-number projection \cite{Morita_JPSJ_2015} and
ground-state refinement methods \cite{Iqbal_PRB_2016}, DMRG suggesting a
gapped one \cite{Zhu_PRB_2015,Hu_PRB_2015}, and the coupled-cluster method
being unable to comment \cite{Li_PRB_2015}. Because studies of the QSL
ground state of the kagome lattice indicate that VMC has a generic preference
for gapless phases and DMRG for gapped ones, there is little to learn from
these results. The problem with this interpretation in Ba$_8$CoNb$_6$O$_{24}$
is that a $J_2$ interaction requires a Co-O-(Nb-)O-O-(Nb-)O-Co path, whose
shortest realization passes through one NbO$_6$ octahedron below the plane
and one above it, via another in-plane CoO$_6$ octahedron [Fig.~\ref{1}(b)].
An alternative path with both intermediate NbO$_6$ octahedra on the same
side of the Co plane is expected to be very similar in strength to the
third-neighbor interaction. However, it appears quite unlikely on
quantum-chemistry grounds that such bonds could have a strength only
one order of magnitude below that of the near-neighbor ($J$) bond.

\section{Summary}

We have performed experimental measurements of the thermodynamic, NQR, and
NMR response of a purely 2D $S = 1/2$ TLAFM. This model system is realized
to high accuracy in the compound Ba$_8$CoNb$_6$O$_{24}$, where the very large
separation of magnetic layers precludes any 3D coupling. With such an ideal
material, it is possible to illustrate clearly the Mermin-Wagner theorem for
a 2D system with continuous spin symmetry by the absence of magnetic ordering
at any temperature and an incipient zero-temperature order reflected in the
increase of the spin-lattice relaxation rate below 0.2 K.

Another feature illustrated by our ideal material is that the thermodynamic
energy scales are anomalously low by comparison with the characteristic
energy scale ($J$) and with unfrustrated systems. While this ``anomaly'' is
in fact well understood in terms of the flattening and suppression of the
one-triplet excitation bands due to frustration, our measurements benchmark
very precisely the efficacy of advanced theoretical and numerical approaches
in reproducing these properties of the TLAFM.

We have measured the complete $(H,T)$ phase diagram of Ba$_8$CoNb$_6$O$_{24}$,
which presents a quantum-corrected version of the classical phase diagram
known from Monte Carlo simulations. It has simply not been possible to
measure this on a TLAFM before now, because no properly 2D system, of
any spin $S$, was known. Further, no analytical or numerical method is
able to compute the quantum phase diagram. We have found four qualitative
corrections, namely the upward shift in the field, $H_{\rm max}$, at which
the maximum transition temperature, $T_m$, occurs, the line of finite slope
separating the field-induced ordered phase from the thermally disordered
regime at low $H$ and $T$, the square-root approach of $T_m$ to zero at
saturation, and the possibility that the only ordered phase is a version
of the up-up-down configuration. All of these changes can be related to the
strong preference of quantum fluctuations for collinear spin states, and the
consequent stabilization of the up-up-down phase.

Finally, our results contain one major exception to the rules of the
$S = 1/2$ Heisenberg TLAFM, namely the magnetically disordered
phase appearing at low $T$ and $H$. In the absence of any evidence for
sample disorder, and of convincing qualitative and quantitative arguments
in favor of a composite spin state arising due to nuclear-spin coupling,
there is a strong possibility that this phase is intrinsic. We believe on
the basis of TLAFM studies to date that this state is not a fifth and
most significant quantum correction to the pure model, but is driven by an
additional term in the spin Hamiltonian, on an energy scale of 0.1 K. The
Ba$_8$CoNb$_6$O$_{24}$ system is sufficiently ideal that most of the potential
candidate anisotropies are excluded. Thus the search for the origin of this
term requires a more detailed analysis than is possible here, and we can
state only that next-neighbor Heisenberg interactions are not ruled out.

In conclusion, Ba$_8$CoNb$_6$O$_{24}$ constitutes a model material for
characterizing the interplay of geometrical frustration, quantum, and
thermal fluctuations in the $S = 1/2$ TLAFM, over a broad range of applied
fields and temperatures down to $T \simeq J/20$. In addition it contains a
surprise at the lowest temperatures, in the form of a possible gapped or
gapless quantum spin liquid, whose origin provides a further challenge to
theory.

\section*{Acknowledgements}

We thank A. Honecker, J. Richter, R. Yu, and Y. Zhou for helpful discussions.
This work was supported by the Ministry of Science and Technology of China
(Grant Nos.~2016YFA0300503 2016YFA0300504, and 2017YFA0302901),
the National Science Foundation of China (Grant Nos.~11222433, 11374364, 
and 11474331), and the Fundamental Research Funds for the Central Universities
and the Research Funds of Renmin University of China (Grant No.~14XNLF08).

\end{document}